\documentclass[english]{IWMS}
\usepackage{graphicx} 
\RequirePackage{amsthm,amsmath}
\newcommand{\ba}{\begin{array}}
\newcommand{\ea}{\end{array}}
\newcommand{\bt}{\begin{tabular}}
\newcommand{\et}{\end{tabular}}
\newcommand{\btb}{\begin{table}}
\newcommand{\etb}{\end{table}}
\newcommand{\bc}{\begin{center}}
\newcommand{\ec}{\end{center}}
\newcommand{\bea}{\begin{eqnarray}}
\newcommand{\eea}{\end{eqnarray}}
\newcommand{\Bea}{\begin{eqnarray*}}
\newcommand{\Eea}{\end{eqnarray*}}
\newcommand{\beq}{\begin{equation}}
\newcommand{\eeq}{\end{equation}}

\newcommand{\bM}{\boldsymbol{M}}
\newcommand{\bSigma}{\boldsymbol{\Sigma}}
\newcommand{\bsigma}{\boldsymbol{\sigma}}

\newcommand{\bxi}{\boldsymbol{\xi}}
\newcommand{\bV}{\boldsymbol{V}}
\newcommand{\bS}{\boldsymbol{S}}

\newcommand{\bs}{\boldsymbol{s}}
\newcommand{\bv}{\boldsymbol{v}}

\newcommand{\bQ}{\boldsymbol{Q}}
\newcommand{\bA}{\boldsymbol{A}}

\newcommand{\bZ}{\boldsymbol{Z}}

\begin{document}

\begin{englishtitle}

\title{Bayesian Covariance Modelling of Large Tensor-Variate Data Sets $\&$ Inverse Non-parametric Learning of the Unknown Model Parameter Vector}{}

\author[1]{Wang, Kangrui}{ Email: kw202$@$le.ac.uk}

\author[1]{Chakrabarty, Dalia}{ Email: dc252$@$le.ac.uk}

\address[1]{Department of Mathematics, University of Leicester}

\maketitle

\begin{abstract}
Tensor-valued data are being encountered increasingly more commonly,
in the biological, natural as well as the social sciences. The
learning of the unknown model parameter vector given such data,
involves covariance modelling of such data, though this can be
difficult owing to the high-dimensional nature of the data, where the
numerical challenge of such modelling can only be compounded by the
largeness of the available data set. Assuming such data to be modelled
using a correspondingly high-dimensional Gaussian Process (${\cal
  GP}$), the joint density of a finite set of such data sets is then a
tensor normal distribution, with density parametrised by a mean tensor
$\bM$ (that is of the same dimensionality as the $k$-tensor valued
observable), and the $k$ covariance matrices
$\bSigma_1,...,\bSigma_k$. When aiming to model the covariance
structure of the data, we need to estimate/learn
$\{\bSigma_1...\bSigma_k \}$ and $\bM$, given tha data. We present a
new method in which we perform such covariance modelling by first
expressing the probability density of the available data sets as
tensor-normal. We then invoke appropriate priors on these unknown
parameters and express the posterior of the unknowns given the
data. We sample from this posterior using an appropriate variant of
Metropolis Hastings. Since the classical MCMC is time and resource
intensive in high-dimensional state spaces, we use an efficient
variant of the Metropolis-Hastings algorithm--Transformation based
MCMC--employed to perform efficient sampling from a high-dimensional
state space. Once we perform the covariance modelling of such a data
set, we will learn the unknown model parameter vector at which a
measured (or test) data set has been obtained, given the already
modelled data (training data), augmented by the test data.

\end{abstract}

\end{englishtitle}

\makecollection



\section{Introduction}
\noindent
Let the causal relationship between observable $\bV$ and model parameter $\bS$ be defined as $\bV=\bxi(\bS)$, where $\bV$ is tensor-variate: $\bV\in{\mathbb R}^{m_1\times m_2... \times m_k}$ . We want to estimate value $\bs^{(test)}$ of $\bS$ at which test data ${\bf D}^{(test)}$--i.e. measured value(s) of $\bV$--is (are) realised. To do this, we need to learn function $\bxi(\cdot)$, which in this case is a tensor-variate function--of the model parameter vector $\bS\in{\mathbb R}^d$. In the presence of training data ${\bf D}$, such supervised learning can be possible by fitting known parametric forms (such as splines/wavelets) to the training data to learn the form of $\bxi(\cdot)$, which can thereafter be inverted and operated upon the test data to yield $\bs^{(test)}$. Here, training data ${\bf D}$ is this set of $n$ values of $\bV$, each generated at a design point, i.e. a chosen value $\bs^{(*)}$ of $\bS$. Thus, ${\bf D}:=\{(\bv_1,\bs_1^{(*)}),\ldots,(\bv_n,\bs_n^{(*)})\}$. However, fitting with splines/wavelets is inadequate in that it does not capture the correlations between the components of a high-dimensional function; also, the computational complications of such fitting--and particularly of inversion of the learnt $\bxi(\cdot)$--increases rapidly with increase in dimensionality. Thus, we resort to the modelling of this high-dimensional data using a correspondingly high-dimensional Gaussian Process (GP), i.e. a tensor-variate GP. 

\section{Method}
\noindent
Thus, the joint probability distribution of a set of $n$ realisations of the $k-1$-variate $\bV$ is a $k$-variate normal distribution with mean $\bM$ and $k$ covariance matrices:
\begin{equation}
p(\bV|\bM,\bSigma_1,...,\bSigma_K) \propto \exp(-\Vert (\bV-\bM)\times_1 \bA_1^{-1} \times_2 \bA_2^{-1} ... \times_k \bA_k^{-1} \Vert^2/2)
\label{eqn:eqn1}
\end{equation}
where the covariance matrix $\bSigma_p = \bA_p \bA^{ T}_p$,$p=1,...,k$. Tensor-variate normal distribution is extensively discussed in the literature, (Xu, Yan $\&$ Yuan \cite{FW97}; Hoff \cite{FW94})

Equation~\ref{eqn:eqn1} implies that the likelihood of $n$ values ($\bv_1,\bv_2,\ldots,\bv_n$) of $\bV$  given the unknown tensor-variate parameters of the GP is $k$-tensor variate normal. So, we write this likelihood and thereafter the posterior probability density of these unknown tensor-variate parameters given the training data (subsequent to the invoking of the priors on each unknown). Once this is achieved, we then sample from the posterior using an MCMC technique to achieve marginal density distributions of each uknown. The learning of $\bs^{(test)}$ could be undertaken by writing the posterior predictive distribution of $\bS$ given the test data $\bv^{(test)}$, and given the tensor-variate parameters learnt using the training data. However, we decide to write the joint posterior probability density of $\bs^{(test)}$ and all the other tensor-variate parameters given training+test data, and sample from this density to obtain the marginals of all the unknowns.

The first step is to write the likelihood of ${\bf D}$ given the tensor-variate mean and covariance matrices of the GP. Here the mean matrix is $\bM \in R^{ m_1 \times
  m_2...\times m_k}$. It may be possible to estimate the mean as a function of $s$ and be removed from the non-zero mean model. Under these circumstance, a general method of estimation, like maximum likelihood estimation or least square estimation, can be used. Then, the Gaussian Process can be converted into a zero mean GP. However, if necessary, the mean tensor itself can be regarded as a random variable and learnt from the data \cite{FW96}. The modelling of the covariance structure of this GP is discussed in the following subsection.

\subsection{Covariance structure}
\noindent
In this context, it is
relevant that a $k$-dimensional random tensor $\bSigma \in R^{ m_1 \times
  m_2...\times m_k}$ can be decomposed to a unit random $k$ dimensional tensor ($\bZ$) and $k$ number of covariance matrix by  \textbf{Tucker product} \cite{FW94}:
  \begin{equation}
   \bSigma = \bZ \times_1 \bSigma_1 \times_2 \bSigma_2 ...  \times_k  \bSigma_k
  \label{eqn:eqn2}
\end{equation}
where the $p$-th covariance matrix is $m_p \times m_p$ matrix and $m_p\in{\mathbb Z}_{> 0}$,  $m_p\in\{m_1, m_2,\ldots,m_k\}$ for the tensor $\bSigma$ that is $m_1\times m_2\times\ldots\times m_k$-dimensional .

 We choose to model the covariance
structure of the GP with a Squared Exponential (SQE) covariance
function. The implementation of this can be expressed in different
ways, but in this initial phase of the project, we perform
parametrisation of the covariance structure using the Tucker Product
that has been extensively studied\cite{FW94}.  It is
recalled that the SQE form can be expressed as 
  \begin{equation}
p(\bV|\bM,\bSigma_1,...,\bSigma_k)=(2\pi)^{-m/2}(\prod_{i=1}^{k}|\bSigma_i|^{-m/2m_i})\times \exp(-\Vert (\bV-\bM)\times_1 \bA_1^{-1} \times_2 \bA_2^{-1} ... \times_k \bA_k^{-1} \Vert^2/2)
  \label{eqn:eqn3}
\end{equation}
where $m=\prod_{i=1}^{k} m_i$ and $\bSigma_p = \bA_p \bA^{ T}_p$.

Although this probability density function is well structured and can in principle be used to model high dimensional data, the computational complicity increases with large and/or high-dimensional data sets. If we do not implement a particular parametric model for the covariance kernels but aim to learn each element of each covariance matrix, the total number of parameters in the covariance structure to be then learnt, ends up as $\displaystyle{\sum\limits^{k}_{p=1} m_p^2}$. This could be a big number for a large data set and the computational demand on such learning can be formidable. Also, the computational task of inverting the covariance matrix $\bSigma_p$ is in itself highly resource intensive, with the demand on time and computational resources increasing with the dimensions of $\bSigma_p$, $p=1,\ldots,k$. 

\section{Application}
\noindent
We perform an empirical illustration of our method, to first learn the
covariance structure of a large astronomical training data set, and
thereafter, employ such learning towards the prediction of the value
of the unknown model parameter at which the test data is realised.
The training data comprises has 216 observations, where an observation
constitutes a sequence of 2-dimensional vectors. In fact, each such
2-dimensional vector is a 2-dimensional velocity vector of a star that
is a neighbour of the Sun, as tracked within an astronomical
simulation \cite{FW95}
of the disk of our Galaxy. There are 50 stars tracked at each design
point i.e. at each assigned value of the unknown model parameter
vector, that is in this application is the location of the Sun in the
two-dimensional, (by assumption), Milky Way disk. In other words,
$\bS$ itself is a 2-dimensional vector. There are 216 design points
used to generate this (simulated) training data that then constitutes
216 number of 50$\times$2-dimensional velocity matrices, with each
velocity matrix generated at each of the 216 design points in this
training data. Thus, the training data ${\bf D}$ in this application
is $216 \times 50 \times 2$-dimensional 3-tensor

To reduce the difficulty of MCMC algorithm, the mean tensor is estimated by the maximum likelihood estimation. 

When building the covariance structure of this training data set, the
likelihood of which is now 3-tensor-normal, we consider three
covariance matrices. Of these, the $216\times216$-dimensional covariance matrix $\bSigma_1$ bears information about the correlation 
between velocity matrices generated at the 216 different values of $\bS$, i.e. at the 216 different solar locations in the Milky Way disk. The $50\times50$ covariance matrix $\bSigma_2$ illustrates the
correlation between any pair of the 50 stars at a given $\bs$, that are tracked in the astronomical simulation and the last covariance matrix $\bSigma_3$ represents the
correlation between the 2 components of the velocity vector of a star that is tracked at a given $\bs$ for its velocty in the astronomical simulation. If we learn the elements of each covarianc matrix directly, we will
have $216\times216+50\times50+2\times2$ number of parameters to learn,
which is too many given limits of time and computational resourse. Thus, we model the covariance kernels using known forms, the parameters of which we then learn from the data. 

In particular, we use the Squared Exponential (SQE)
covariance function to model the $216\times216$ matrix $\bSigma_1$ and learn the correlation lengths--or rather their reciprocals, the smoothing parameters--using the training data. As the 216 velocity matrices are each generated at a respective value of $\bS$, $\bSigma_1$ can be written as $\bSigma_1=[a_{ij}]$ where $i,j=1,\ldots,216$ with
$$a_{ij}=\displaystyle{\exp\left[-\left(\bs_i- \bs_j\right)^T \bQ_1 \left(\bs_i -\bs_j\right)\right]},$$
where $\bQ_1$ is a $d\times d$ square diagonal matrix, with $\bS\in{\mathbb R}^d$. As $d=2$ in our application, we learn 2 smoothness parameters. 

The covariance matrix $\bSigma_2$ quantifies correlation amongst the different stellar velocity vectors generated at a given $\bs$. The 50 stellar velocity vectors that are recorded at a given $\bs$ are chosen over other values of stellar velocity vectors. Given that the velocity vector of each star is 2-dimensional, we again learn 2 smoothness parameters (diagonal elements of matrix $\bQ_2$), using an SQE model. 

In addition we learn the 4 parameters of the covariance matrix $\bSigma_3$.

Thus, we will have 8 parameters ($q_{11}^{(1)}$,$q_{22}^{(1)}$,$q_{11}^{(2)}$,$q_{22}^{(2)}$,$\sigma_{11}^{(3)}$,$\sigma_{12}^{(3)}$,$\sigma_{21}^{(3)}$,$\sigma_{22}^{(3)}$) of the covariance structure to learn from the data, where these parametersare defined as in: 
$$\bQ_1= \begin{pmatrix}
  q^{(1)}_{11} & 0 \\
0 & q^{(1)}_{22} \\
\end{pmatrix}  ; 
\bQ_2= \begin{pmatrix}
  q^{(2)}_{11} & 0 \\
0 & q^{(2)}_{22} \\
\end{pmatrix} ; 
\bSigma_3= \begin{pmatrix}
  \sigma^{(3)}_{11} & \sigma^{(3)}_{12} \\
\sigma^{(3)}_{21} & \sigma^{(3)}_{22} \\
\end{pmatrix}$$ In the initial phase of the project that is currently underway, we write the joint posterior probability density of the unknown parameters and sample from it using a variant of the metropolis-Hastings algorithm, referred to as Transformation-based MCMC (TMCMC). To write the posterior, we impose uniform priors on each of our unknowns. 
\begin{table}[ht] \bc \caption{Priors for parameters \label{tb2}}
\smallskip {\small 
\tabcolsep=10pt \bt{*{7}{c}} \hline {Parameters} &{ }&{Prior}
\\\hline
$q_{11}^{(1)}$  &Uniform  & $\pi(q_{11}^{(1)})\propto 1$ \\
$q_{22}^{(1)}$ &Uniform & $\pi(q_{22}^{(1)})\propto 1$\\
$q_{11}^{(2)}$ &Uniform & $\pi(q_{11}^{(2)})\propto 1$\\
$q_{22}^{(2)}$ &Uniform & $\pi(q_{22}^{(2)})\propto 1$\\
$\bSigma_3$ &Non-informative & $\pi(\bSigma_3)\propto {\vert \bSigma_3 \vert}^{-1/2}$  \\\hline

\et} \ec \end{table}

The proposal density that we use in our MCMC scheme, to generate updates for each of our parameters is tabulated within the section in which TMCMC is described. The results of our learning and estimation of the mean and covariance structure of the GP used to model this tensor-variate data, is discussed below in Section~\ref{sec:results}. Once this phase of the work is over, we will proceed to include the test and training data both, to write the joint posterior probability density of $\bs^{(test)}$ and the 8 unknowns in $\bQ_1,\bQ_2,\bSigma_3$, and learn all these parameters.

\section{Transformation based MCMC}
\noindent
We are using the Transformation based MCMC algorithm to estimating the parameters. Although the TMCMC method will lose some of the information, the method is efficient in high dimensional distributions.

	\begin{itemize}
					\item 1.Set initial value $s_0,q_0^{(1)},\ldots,q_0^{(k)}$ , counter $n=1$ and a forward probability $p_0,\ldots,p_k$ 
					\item 2.Generate $e\sim{ Gamma}(1,1)$ and $u\sim{\cal U}(0,1)$ independently.
					\item 3.If $u<p_0$, let $s'=s_{n-1}+\beta_0e$. Else, let $s'=s_{n-1}-\beta_0e$
					\item 4.Repeat step 2 and step 3 for $q'_1,\ldots,q'_k$.
					\item 5.Calculate the acceptance rate:
					$$\alpha=\frac{\prod_{i\in D}p_i\times \prod_{j\in D^c}(1-p_j)}{\prod_{i\in D}(1-p_i)\times \prod_{j\in D^c}p_j}\times \frac{posterior(s', q'_1,...q'_k)}{posterior(s_{n-1}, q_{n-1}^{(1)},...q_{n-1}^{(k)})}$$
					where, set $D$ is the elements which has the backward transform($u>p$) and set $D^c$ is the elements which has the forward transform($u<=p$).
					\item 6.Accept $s',q'_1,\ldots,q'_k$ as $s_{n},q_{n}^{(1)},...,q_n^{(k)}$ with probability $\alpha$ or drop $s',q'_1,\ldots,q'_k$ with probability $1-\alpha$ 
					\item 7.Repeat 2 to 6 until the chain get convergence.
					\end{itemize}

\section{Results}
\label{sec:results}

\begin{figure}[htbp]
\begin{minipage}[t]{0.5\textwidth}
\centering
\includegraphics[trim=100 250 100 250,scale=0.5]{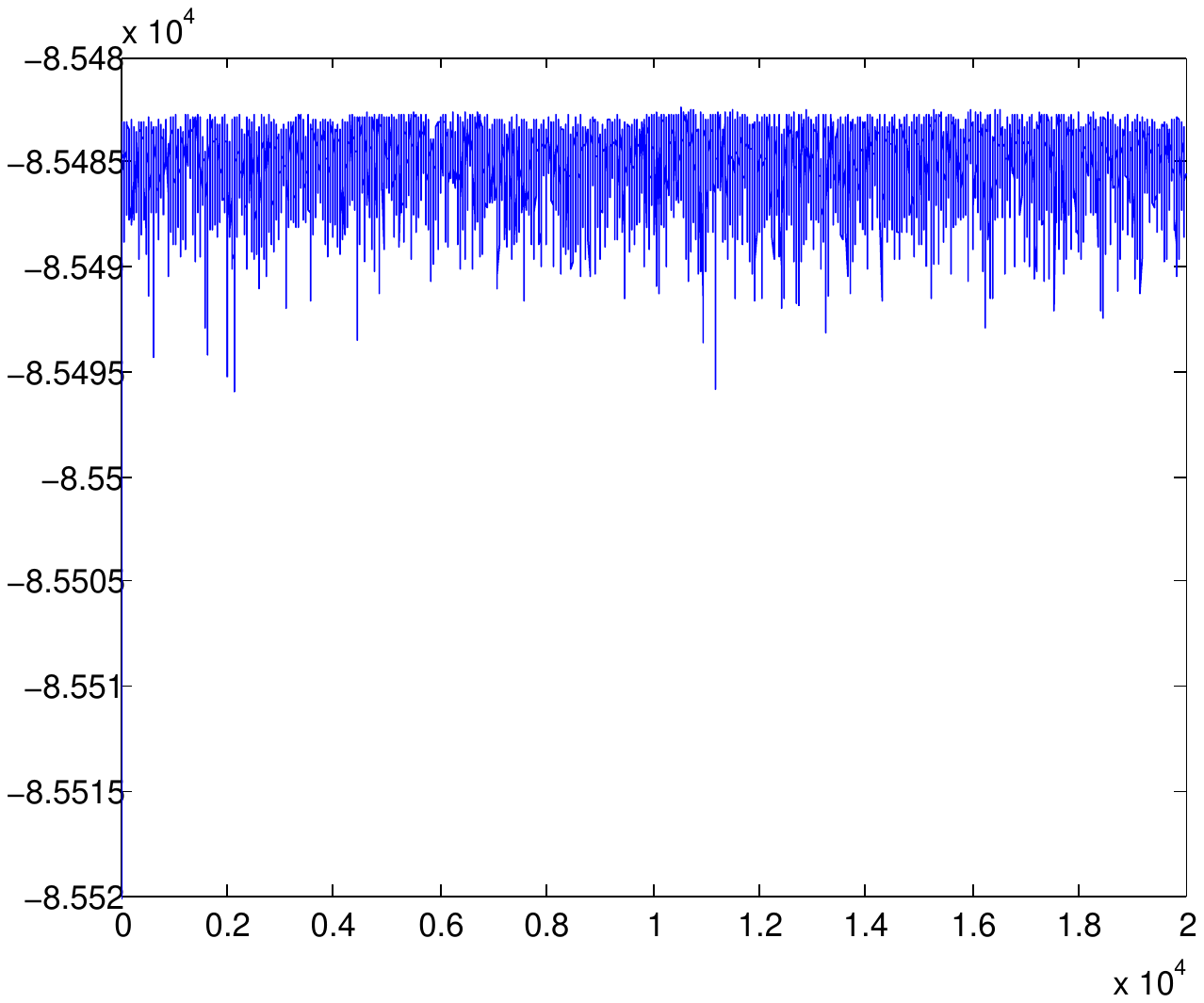}
\caption{trace of the likelihood generated by TMCMC} \label{figure 1}
\end{minipage}
\begin{minipage}[t]{0.5\textwidth}
\includegraphics[trim=100 250 100 250,scale=0.5]{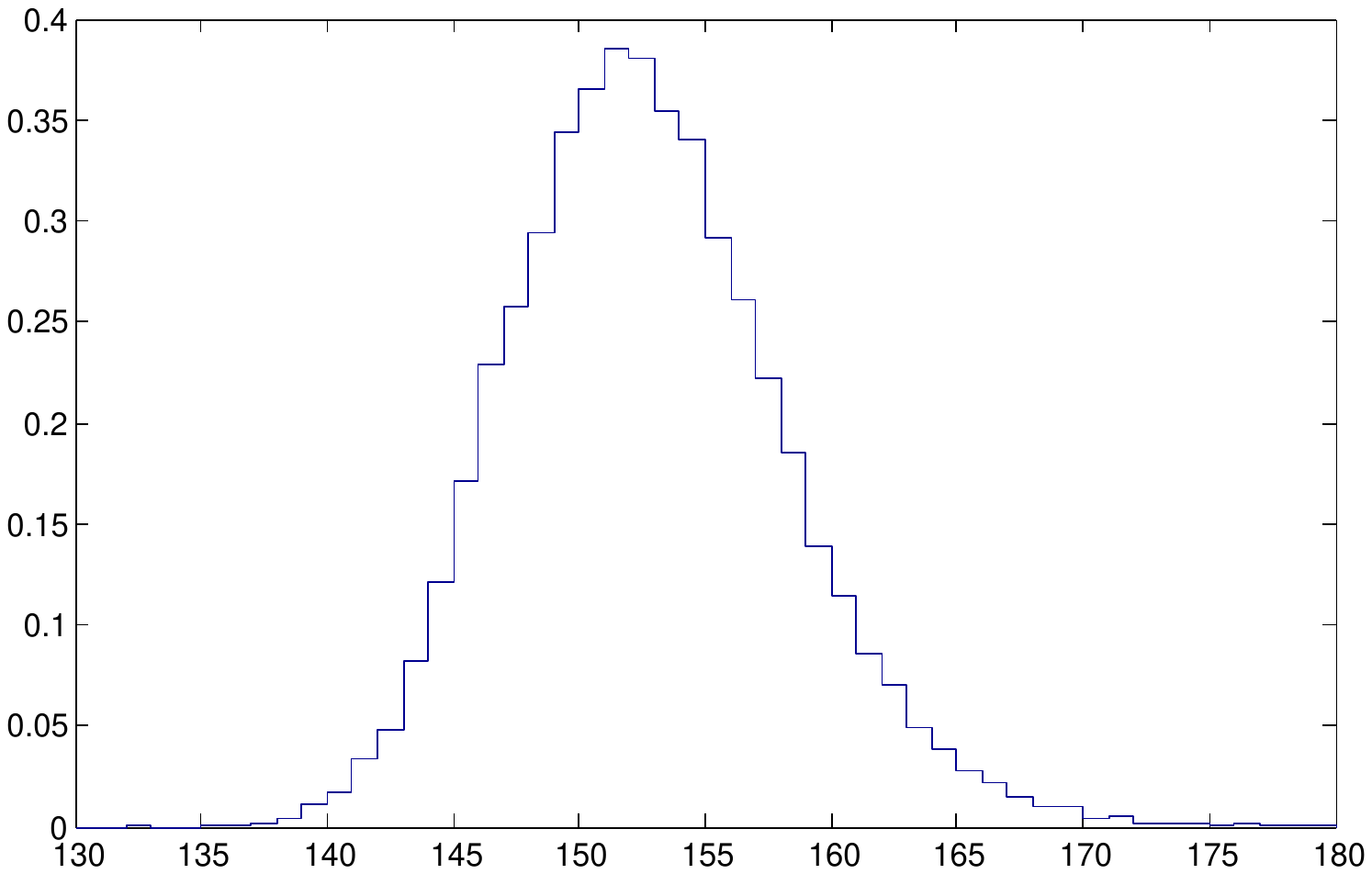}
\caption{marginal probability density for $q^{(1)}_{11}$} \label{figure 2}
\end{minipage}
\end{figure}
\begin{figure}[htbp]
\begin{minipage}[t]{0.5\textwidth}
\centering
\includegraphics[trim=100 250 50 250,scale=0.5]{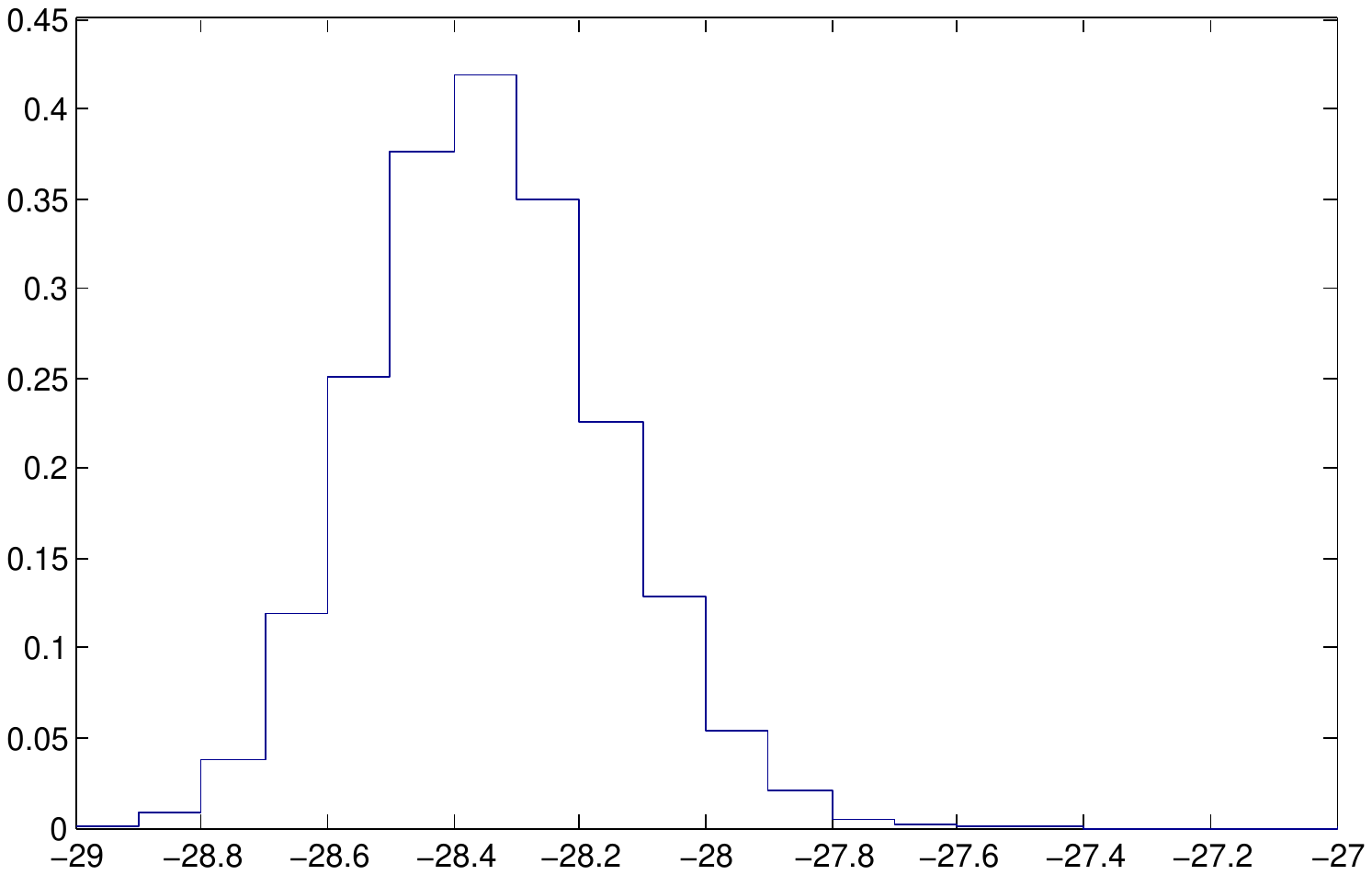}
\caption{marginal probability density for $\bsigma^{(3)}_{11}$}\label{figure 3} 
\end{minipage}
\begin{minipage}[t]{0.5\textwidth}
\includegraphics[trim=80 250 50 250,scale=0.5]{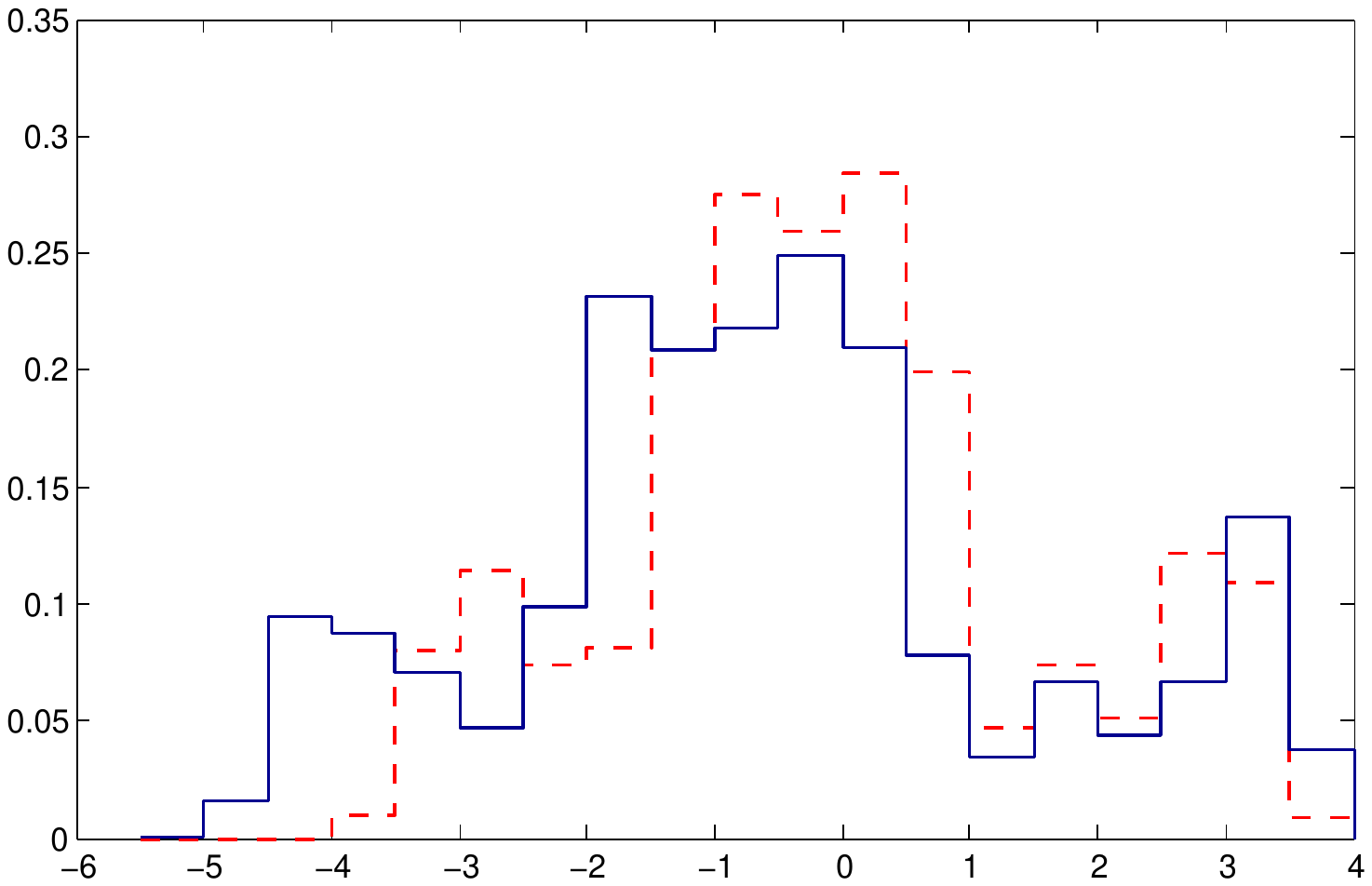}
\caption{marginal probability density for $\bsigma^{(3)}_{12}$ in full line and $\bsigma^{(3)}_{21}$ in broken line} \label{figure 4}
\end{minipage}
\end{figure}

In the top left panel of Figure~\ref{figure 1}, we present the
trace of the likelihood of the training data given the 8 unknowns in $\bQ_1,\bQ_2,\bSigma_3$,
with $2\times 10^4$ of iterations. The stationarity of the trace betrays the
achievement of convergence of the chain.

The marginal posterior probability densities of each unknown parameter
is also learnt using TMCMC. The same for parameters $q^{(1)}_{11}$ (Figure~\ref{figure 2}), $\bsigma_{11}$ (Figure~\ref{figure 3}) and
$\bsigma_{12}$ (Figure~\ref{figure 4})are shown in the top right and bottom left and right panels. As noticed in the inequality of the marginals of the non-diagonal elements of $\bSigma_3$ shown in the bottom panels of this figure, the covariance structure for this astronomical data set does not appear to adhere to stationarity. Had the covariance been stationary, the $1,2$-th and $2,1$-th elements would be equal, i.e. their marginals would coincide. But such is not the case as evident from comparing the two density in figure~\ref{figure 4} which shows a drift from $\bsigma_{12} $ to $\bsigma_{21}$. This further suggests that our modelling of the $\bSigma_2$ matrix using SQE covariance function is pre-matured. We are exploring the implementation of non-stationary covariance modelling of $\bs$.

\end{document}